**Formation of Single-mode Laser in Perovskite Nanowire via Nano-manipulation**

*Kaiyang Wang, Zhiyuan Gu, Shuai Liu, Jiankai Li, Shumin Xiao\*, and Qinghai Song\**

Dr. K. Y. Wang, Dr. Z. Y. Gu, Dr. S. Liu, Mr. J. K. Li, Prof. Q. H. Song
Integrated Nanoscience Lab, Department of Electrical and Information Engineering, Harbin Institute of Technology, Shenzhen, 518055, China
E-mail: qinghai.song@hitsz.edu.cn
Prof. S. M. Xiao
Integrated Nanoscience Lab, Department of Material Science and Engineering, Harbin Institute of Technology, Shenzhen, 518055, China
E-mail: Shumin.xiao@hitsz.edu.cn

Perovskite based micro- and nano- lasers have attracted considerable research attention in past two years. However, the properties of perovskite devices are mostly fixed once they are synthesized. Here we demonstrate the tailoring of lasing properties of perovskite nanowire lasers via nano-manipulation. By utilizing a tungsten probe, one nanowire has been lifted from the wafer and re-positioned its two ends on two nearby perovskite blocks. Consequently, the conventional Fabry-Perot lasers are completely suppressed and a single laser peak has been observed. The corresponding numerical model reveals that the single-mode lasing operation is formed by the whispering gallery mode in the transverse plane of perovskite nanowire. Our research provides a simple way to tailor the properties of nanowire and it will be essential for the applications of perovskite optoelectronics.

## 1. Introduction

Organic-inorganic hybrid perovskites have been intensively studied in past few years due to their potential applications in low-cost, high-efficient light harvesting [1-5]. Although the perovskites are usually synthesized from the solution phase at low temperature, they have already shown exceptional material properties for high performance solar cells, e.g. low defect densities, high linear and nonlinear absorption coefficients, long electron/hole diffusion lengths, and larger carrier mobility [2, 3]. In 2015, a record high light convention coefficient (20.1%) has been experimentally demonstrated in lead halide perovskite CH3NH3PbI3 [5]. In additional to the efficient light harvesting, the long optical and electrical transports of perovskites have also promised their practical applications in optoelectronic devices [2, 6]. As the optical properties of perovskites are strongly dependent on the crystal qualities, perovskites with reduced dimensions such as hexagon shaped and rectangle shaped microplates have been rapidly synthesized and the whispering gallery modes (WGM) based lasing actions have been reported [7, 8]. Very recently, perovskite nanowires have also attracted large research attention [9-11]. Based on the higher crystal quality and longer carrier diffusion length of perovskite nanowire, Zhu et al have successfully demonstrated the nanowire lasers with high quality (Q) factors and record low thresholds [9]. Very recently, low-threshold nanowire lasers under two-photon optical excitation have also been realized [10]. The measured threshold can even be comparable to that of microplate lasers under single-photon pumping. All these achievements have clearly shown the bright futures of perovskites in both linear and nonlinear optoelectronic devices.

Compared with two-dimensional microplates, one-dimensional nanowires have several additional advantages. In the transverse plane, the sizes of perovskite nanowires are usually on the order of several hundred nanometers, which are small enough to couple with other on-chip integrated circuits or nano-photonic network [12]. In the longitude direction, the nanowires are usually tens of microns in length. Thus they can easily bridge the typical fiber systems with nano-circuits. Mostly interestingly, the nanowires are large enough to be observed under optical microscope and thus the positions and shapes can be easily tailored by the well-known nano-manipulation [12, 13]. Such kind of manipulation can simply modify the corresponding lasing actions inside nanowires. And it is also a key step to integrate the bottom-up synthesized devices with top-down fabricated photonic circuits. In this research, we tailor the perovskite nanowires via nano-manipulation and study their lasing behaviors. Different from the Fabry-Perot lasing actions in typical perovskite nanowires, we have successfully observed single-mode lasing actions within the transverse plane of perovskite nanowire by repositioning it on the gap between two nearby perovskites.

## 2. Results and Discussions

The perovskite nanowires have been simply synthesized with one-step solution based self-assemble method, which has been reported by Liao et al recently [8]. Basically, equal volumes of solutions of $CH_3NH_3Br$ (0.2M) and $PbBr_2$ (0.2M) in DMF (N,N-dimethylformamide) were mixed, yielding stock solution 1 of $CH_3NH_3Br \cdot PbBr_2$ (0.1M). Then the stock solution was dip-casted onto a glass slide, which was amounted on a teflon stage in a beaker containing dichloromethane (DCM) leveled below the Teflon stage. The beaker was sealed with a porous Parafilm (3M) to control the evaporation rate of solution 1. Diffusion of DCM vapor into solution 1 induced the nucleation and subsequent growth of perovskite nanowires.

We then carried out the lasing experiments by pumping the as-synthesized perovskite nanowire. The samples were placed onto a translation stage under the microscope and excited by a frequency-doubled laser (400 nm, using a BBO crystal) from a regenerative amplifier (repetition rate 1 kHz, pulse width 100 fs, seeded by MaiTai, Spectra Physics). The pump light was focused through a 40x objective lens and the beam size was adjusted to ~ 100 micron, which was large enough to uniformly pump the nanowires. The emitted lights were collected by the same objective lens and coupled to a CCD (Princeton Instruments, PIXIS UV enhanced CCD) coupled spectrometer (Acton SpectroPro s2700) via a multimode fiber. **Figure 1** summarizes the lasing behaviors in a typical perovskite nanowire. The top-view scanning electron microscope (SEM) image of a typical $CH_3NH_3PbBr_3$ nanowire is shown in **Figure 1(a)**. We can see that the length and width of the nanowire are 18.15 μm and 717 nm, respectively. Periodic laser peaks have been observed when the nanowire was pumped above lasing threshold (**see Figure 1(b)**). The measured free spectral range (FSR) is about 1.56nm. **Figure 1(c)** shows the evolution of emission spectrum with the increase of pumping power. When the pumping density was below 8.2 μJ/cm$^2$, the recorded spectrum showed a broad peak with full width half maximum (FWHM) around 23 nm. Once the pumping density was above 8.2 μJ/cm$^2$, periodic lasing peaks emerged immediately. Meanwhile, two bright spots could be clearly seen in the fluorescent microscope image (inset in **Figure 1(b)**). The integrated output intensity has been plotted in **Figure 1(d)**, where a laser threshold can be clearly observed. All these characteristics are consistent with previous reports.

This kind of lasing action has been well explored in the literatures [9-11]. As the refractive index of perovskites (n = 2.55) is much larger than the air (n = 1) and glass substrate (n = 1.5), the light can be well trapped by total internal reflection along the waveguide modes. Consequently, the light can be reflected back and forth by two end-facets and form Fabry-Perot modes. Due to the large nanowire length, the FSR is much smaller than the width of gain spectrum [14]. Thus multiple lasing peaks were observed in experiment. We note that multiple lasing peaks (see **Figure 1** for an example) are quite generic in our experiment. While the mode around the peak position of photoluminescence can have the largest gain, the spatial hole burning caused inhomogeneous gain saturation and crystal inhomogeneity will quickly smear out the single-mode operation [15].

Considering the importance of monochromaticity of laser, it is very interesting to explore the possibility to achieve single-mode operation in perovskite. In perovskite nanowire, the sizes in the transverse plane are comparable to the lasing wavelength. Similar to the previous report [16], the light can be easily confined by the rectangle shape and form conventional WGMs in the transverse plane. Meanwhile, the FSR of such kind of WGMs are usually much larger than the spectral range of photoluminescence. Thus the transverse modes are good candidates to achieve single-mode laser. However, the modes in transverse plane are usually hard to be excited and observed in perovskite nanowire because the Fabry-Perot lasers have much longer amplification length. Therefore, suppressing the Fabry-Perot modes along the longitude direction is a key step to observe the transverse direction single-mode laser.

In this research, we suppressed the Fabry-Perot modes by introducing radiation losses of the waveguide modes. As depicted in **Figure 2(a)**, the main part of perovskite nanowire was positioned onto a thick microplate, and the right part was suspended in air. The height and width of perovskite nanowire were 850 nm and 939 nm, respectively (we set these parameters following the experimental data, see below). When the light is propagating in the right part, the fundamental mode is shown as **Figure 2(b)**. We can see that the light is mainly localized within the nanowire. Once the light reaches the left part, the light quickly leaves the nanowire and enters the microplate due to the breaking of total internal reflection at the bottom surface of nanowire [17]. Because two waveguide modes have quite different mode profiles, the convention coefficient at the interface is low and thus huge radiation loss has been introduced to the waveguide [18, 19]. This radiation loss can be easily calculated via overlap integration and clearly seen from **Figure 2(d)**, where the transmission quickly reduces with the refractive index of microplate approaching perovskite. Consequently, if the perovskite nanowire is placed onto a perovskite microplate, the Q factors of Fabry-Perot modes are significantly spoiled and the lasing actions in transverse plane can be expected.

We then numerically calculated the resonances within the transverse plane. The results are summarized in **Figure 3**. In a spectral range between 510 nm – 590 nm, numerous modes have been observed. Most of the Q factors are smaller than 250. Only one resonance at 551 nm (marked as mode-1) has the Q value around 750. As the lasing modes usually correspond to the high Q resonances, **Figure 3(a)** clearly indicates the possibility of single-mode operation in the transverse plane of perovskite nanowire [16]. The formation of high Q mode in Figure 3(a) can be understood as the following. The main resonances in **Figure 3(a)** are still Fabry-Perot modes [20-22]. When two sets of resonances with different FSRs approach each other at particular wavelength, they have the possibility to strongly couple with each other if both of their mode numbers are even or odd in two directions [20, 23]. Then the interference between two resonances can form two hybrid modes with unique field patterns. One hybrid mode has main field distributions away from the corner (**Figure 3(b)**), whereas the other one is mostly confined around four corners (**Figure 3(c)**). Following the previous researches on polygon shaped microcavities, two types of leakages can be identified [20, 24, 25]. One is the boundary wave leakage, which is caused by the scattering of waves at the cavity boundary. The other one is the pesodointegrable leakage, which is caused by the deviation of waves from the close ray orbit. In general, the boundary wave leakage can be simply suppressed by tailoring the field distributions [20, 23]. Therefore, the Q factor of the mode in **Figure 3(b)** can be much larger than the others. Meanwhile, the associated hybrid mode (in **Figure 3(c)**) has much lower Q factor.

As the perovskite nanowire can be clearly seen under optical microscope, it is thus much easier to physically manipulate individual nanowires. Then we experimentally examined our above analysis and numerical calculation. In our experiment, a tungsten probe was fixed onto a three-dimensional translation stage and placed nearby perovskite nanowires under microscope (see the schematic picture of setup in **Figure 4(a)**). In experiments, numerous perovskite nanowires have been synthesized simultaneously. Some nanowires were lying on the substrate and contacted the substrate well (see **Figure 1(a)**). These nanowires can be tailored on the substrate, but they are hard to be lifted from the substrate. We thus selected the nanowires that were standing on the substrate (see example in **Figure 4(b)**). The tungsten probe was gradually pushed to the position of nanowire. Once they contacted directly, the nanowire adhered to the tungsten probe very well and could be easily lifted. **Figures 4(c)** and **4(d)** show the microscope image and florescent microscope image of the picked nanowire. Compared with **Figure 4(b)**, we can see that the background nanowire and microplates are all out of focus. Thus we can confirm that the nanowire has been fully lifted from the substrate.

After lifting the nanowire, we moved the translation stage and placed it onto the desired position. All these manipulation were conducted under optical microscope. **Figure 5(a)** shows an example SEM image of the final structures. We can clearly see that the nanowire was seating onto two perovskite microplates across the gap between them. The width and height of nanowire were 939 nm and 850 nm, which are the same as the settings in **Figure 2**. Following above analysis, we know that dramatic radiation losses have been introduced at the positions where the nanowire touched two microplates. Consequently, the Fabry-Perot modes shall be greatly suppressed and the transverse mode can be observed.

Then we optically excited the nanowire and studied its lasing characteristics to verify the above analysis. All the results are summarized in **Figure 5(c)** and **Figure 5(d)**. A broad emission peak centered at 545nm has been observed. This is a typical spontaneous emission peak with FWHM ~ 33 nm. The intensity of photoluminescence increased slowly with the increase of pumping density at the beginning. Once the pumping density above 5 $\mu J/cm^2$, the intensity increased dramatically. **Figure 5(d)** shows the output intensities as a function of pumping power. A clear "S" shaped curve can be clearly observed, showing transition from spontaneous emission to amplification and finally to the gain saturation. The significant difference between the results in **Figure 5** and **Figure 1** lies in the laser spectrum. Once the pumping density was above 5 $\mu J/cm^2$, a sharp laser peak emerged in the laser spectrum (see **Figure 5(c)**). While single-mode emissions at threshold have been frequently observed in microlasers, here the single-mode laser was maintained in a wide range of pumping density from laser threshold to the gain saturation. Since no parity-time symmetry or partially pumping has been applied and the Fabry-Perot modes have been suppressed, the results in **Figure 5(c)** shall come from the resonances in transverse plane [26-28]. Similar to the numerical calculations, only one resonance has relatively high Q factor due to the mode coupling and the nearby high Q modes are far away from the spectral range. As shown in **Figure 3(c)**, the calculated Q factor of mode-1 is about 750, which is also very close to the experimentally measured Q value at the threshold ($\Delta\lambda \sim 0.83$ nm, giving a $Q \approx \lambda/\Delta\lambda = 667$).

To further confirm the lasing action in transverse plane, we have also studied the fluorescent microscope images. One example is shown in **Figure 5(b)**. Different from the inset in **Figure 1(b)**, bright spots at two ends of the nanowires disappeared even though the pumping density was above the threshold. In contrast, three bright spots can be observed in **Figure 5(b)**. Two main spots appeared at the edges of perovskite nanowire and the third tiny one rose at the crossing line between nanowire and the larger microplate. The latter one is caused by the scattering at the contact region, which is consistent with our numerical calculation in **Figure 2**. Thus the suppression of Fabry-Perot modes along the longitude direction can be confirmed and the lasing action can only be formed within the transverse plane. As mentioned above, both the boundary leakage and the pesudointegrable leakage of boundary mode occur at the corners, which correspond to the edges of perovskite nanowire. Therefore, the two brighter spots in the fluorescent microscope image can be well understood and confirm the single-mode laser in transverse plane well.

**3. Conclusion**

In conclusion, we have studied the lasing actions in perovskite nanowires. By physically manipulating the positions of nanowires, we have successfully implemented radiation losses to the waveguide modes. The conventional Fabry-Perot lasers along the longitude direction have been suppressed and the lasing action within the transverse plane has been observed. Due to the relatively small sizes in transverse direction, single mode laser operation has been achieved within a wide range of pumping power. We believe that our research shall shed light on tailoring the lasing properties of perovskite nanowires and extending their potential applications in integrated photonic circuits.

**Acknowledgements**
This work is supported by NSFC11204055, NSFC61222507, NSFC11374078, NCET-11-0809, Shenzhen Peacock plan under the Nos. KQCX2012080709143322 and KQCX20130627094615410, and Shenzhen Fundamental research projects under the Nos. JCYJ20130329155148184, JCYJ20140417172417110, JCYJ20140417172417096.

Received: ((will be filled in by the editorial staff))
Revised: ((will be filled in by the editorial staff))
Published online: ((will be filled in by the editorial staff))

[1]     G. Xing, N. Mathews, S. Sun, S. S. Lim, Y. M. Lam, M. Grätzel, S. Mhaisalkar, T. C. Sum, *Science* **2013**, *342*, 344.

[2]     G. Xing, N. Mathews, S. S. Lim, N. Yantara, X. Liu, D. Sabba, M. Grätzel, S. Mhaisalkar, T. C. Sum, *Nat. Mater.* **2014**, *13*, 476.

[3]     Q. Dong, Y. Fang, Y. Shao, P. Mulligan, J. Qiu, L. Cao, J. Huang, *Science* **2015**, *347*, 967.

[4]     W. S. Yang, J. H. Noh, N. J. Jeon, Y. C. Kim, S. Ryu, J. Seo, S. I. Seok, *Science* **2015**, *348*, 1234-1237.

[5]     D. Shi, V. Adinolfi, R. Comin, M. J. Yuan, E. Alarousu, A. Buin, Y. Chen, S. Hoogland, A. Rothenberger, K. Katsiev, Y. Losovyj, X. Zhang, P. A. Dowben, O. F. Mohammed, E. H. Sargent, O. M. Bakr, *Science* **2015**, *347*, 519.

[6]     S. T. Ha, X. Liu, Q. Zhang, D. Giovanni, T. C. Sum, Q. Xiong, *Adv. Opt. Mater.* **2014**, *2*, 838.

[7]     Q. Zhang, S. T. Ha, X. Liu, T. C. Sum, Q. Xiong, *Nano Lett.* **2014**, *14*, 5595.

[8]     Q. Liao, K. Hu, H. H. Zhang, X. D. Wang, J. N. Yao, H. B Fu, *Adv. Mater.* **2015**, *27*, 3405.

[9]     H. Zhu, Y. Fu, F. Meng, X. Wu, Z. Gong, Q. Ding, M. V. Gustafsson, M. Tuan Trinh, S. Jin, X-Y. Zhu, *Nat. Mater.* **2015**, *14*, 636.

[10]   Z. Y. Gu, K. Y. Wang, W. Z. Sun, S. Liu, J. K. Li, S. M. Xiao, Q. H. Song, *arXiv*:1510.03987.

[11]   J. Xing, X. F. Liu, Q. Zhang, S. THalf, Y. W. Yuan, T. C. Sum, Q. H. Xiong, *Nano Lett.* **2015**, *15*, 4571.

[12]   P. J. Pauzauskie, P. D. Yang, *Mater. Today* **2006**, *9*, 36.

[13]   P. J. Pauzauskie, D. J. Sirbuly, P. D. Yang, *Phys. Rev. Lett.* **2006**, *96*, 143903.

[14]   Z. Y. Gu, N. Zhang, Q. Lyu, M. Li, S. M. Xiao, Q. H. Song, *arXiv*: 1505.03937.

[15]   L. W. Casperson, *J. Appl. Phys.* **1975**, *46*, 5194.

[16]   Q. H. Song, H. Cao, S. T. Ho, G. S. Solomon, *Appl. Phys. Lett.* **2009**, *94*, 0611109.

[17]   K. Y. Wang, Z. Y. Gu, W. Z. Sun, J. K. Li, S. M. Xiao, Q. H. Song, *ACS Photonics* **2015**, *2*, 1278.

[18]   M. Skorobogatiy, *Nanostructured and Subwavelength Waveguide: Fundamentals and Applications*, Wiley, Hoboken, **2012**.

[19]   Z. Y. Gu, S. Liu, S. Sun, K. Y. Wang, Q. Lyu, S. M. Xiao, Q. H. Song, *Sci. Rep.* **2015**, *5*, 9171.

[20]   J. Wiersig, *Phys. Rev. Lett.* **2006**, *97*, 253901.

[21]   H. Long, Y. Z. Huang, X. W. Ma, Y. D. Yang, J. L. Xiao, L. X. Zou, B. W. Liu, *Opt. Lett.* **2015**, *40*, 3548.

[22]   A. W. Poon, F. Courvoisier, R. K. Chang, *Opt. Lett.* **2001**, *26*, 632.

[23]   Q. H. Song, H. Cao, *Phys. Rev. Lett.* **2010**, *105*, 053902.

[24]   J. Wiersig, *Phys. Rev. A* **2003**, *67*, 023807.

[25]   Q. H. Song, L. Ge, J. Wiersig, H. Cao, *Phys. Rev. A* **2013**, *88*, 023834.

[26]   L. Feng, Z. J. Wong, R. M. Ma, Y. Wang, X. Zhang, *Science*, **2014**, *346*, 972.

[27]   H. Hodaei, M. A. Miri, H. Heinrich, D. N. Christodoulides, M. Khajavikhan, *Science*, **2014**, *346*, 975.

[28]    M. Li, N. Zhang, K. Y. Wang, J. K. Li, S. M. Xiao, Q. H. Song, *Sci. Rep.* **2015**, *5*, 13682.

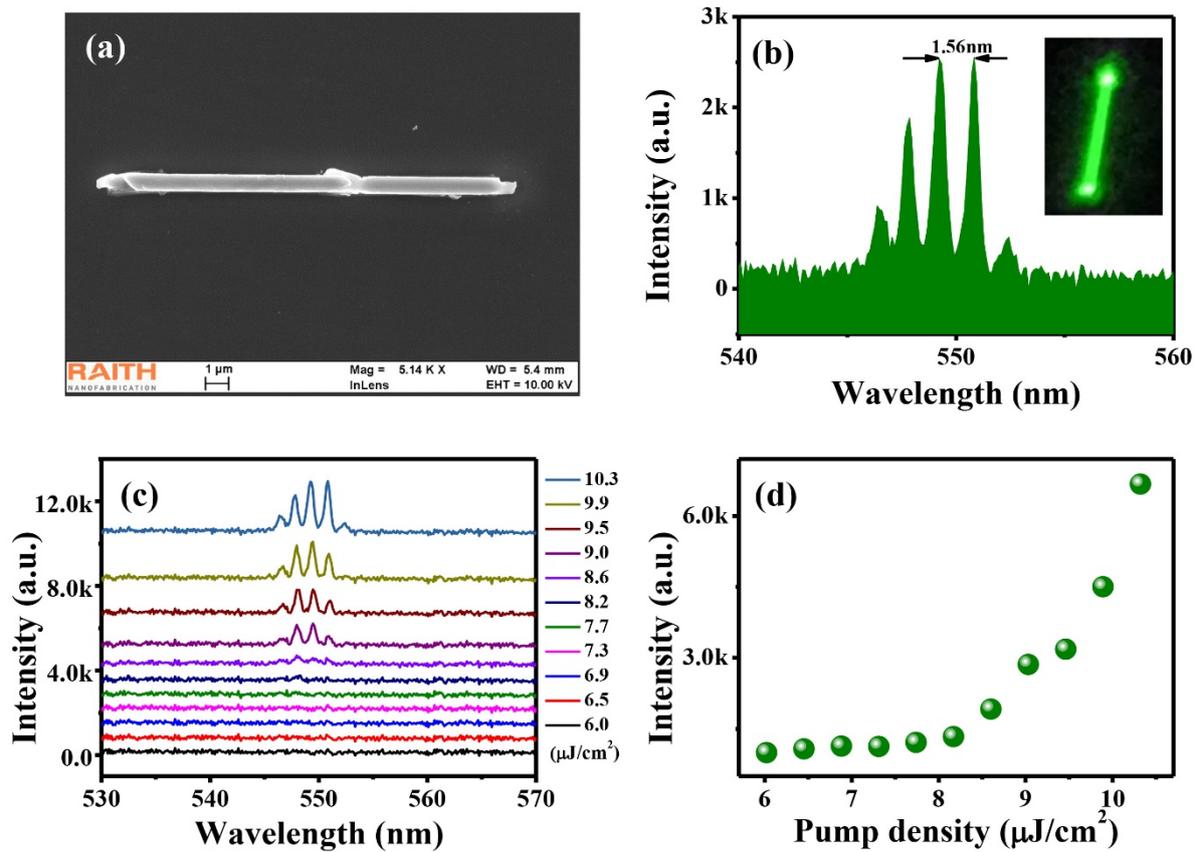

**Figure 1. Typical multimode lasing action in a single perovskite nanowire.** (a) Top-view SEM image of single nanowire of bromide perovskite $CH_3NH_3PbBr_3$. The nanowire is positioned onto a glass substrate. The length and width of nanowires are 18.15 μm and 717 nm, respectively. (b) Multimode lasing spectrum under optical excitation. The inset shows the corresponding fluorescent microscope image. Here the pumping density is 10.3 μJ/cm$^2$. (c) The spectra of perovskite nanowires under different pumping density. (d) The threshold behavior of perovskite nanowire.

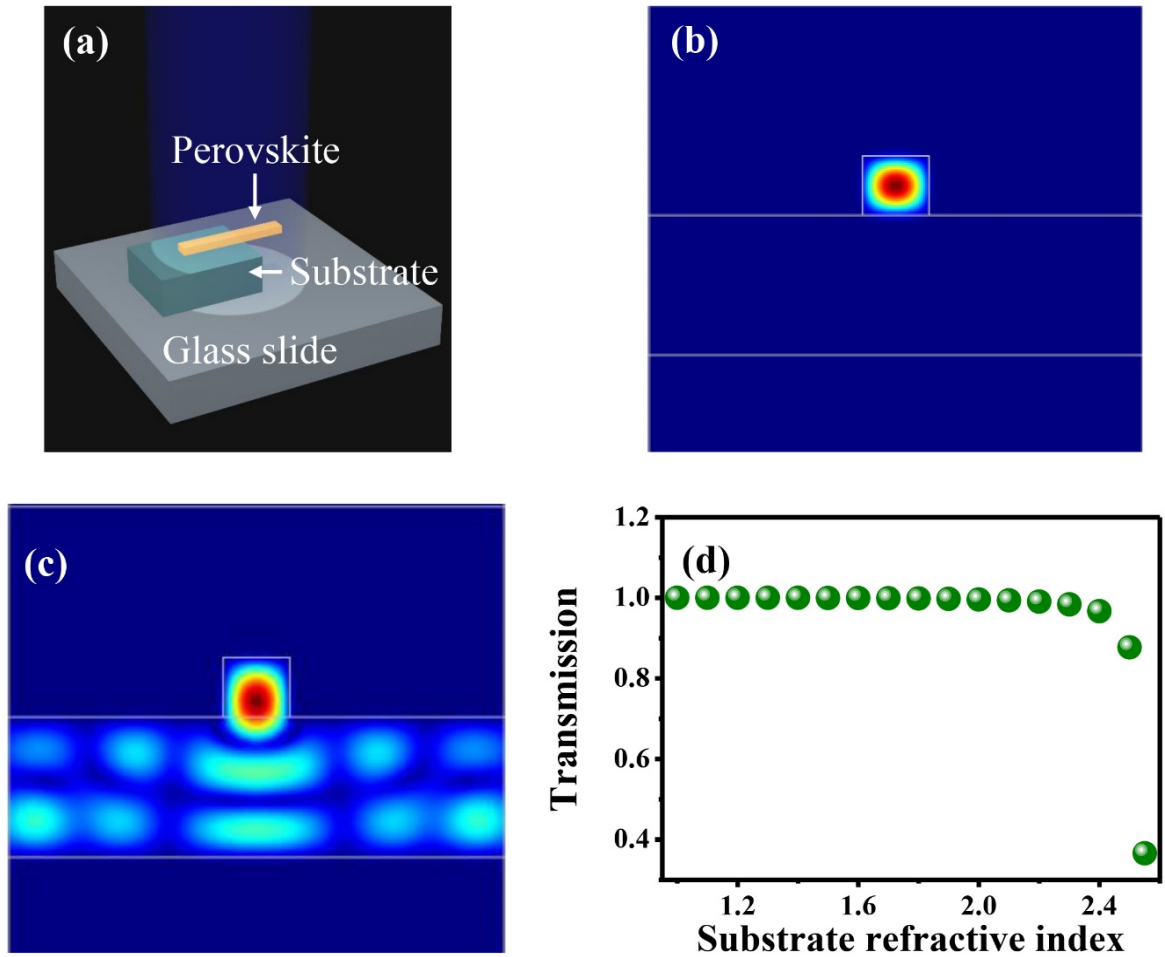

**Figure 2. Introducing the radiation loss to Fabry-Perot cavity modes.** (a) Schematic picture of the structure. (b) and (c) show the field distributions of waveguide modes in the transverse plane. (d) Mode conversion coefficient between the modes in perovskite nanowire with air surrounding and on substrate.

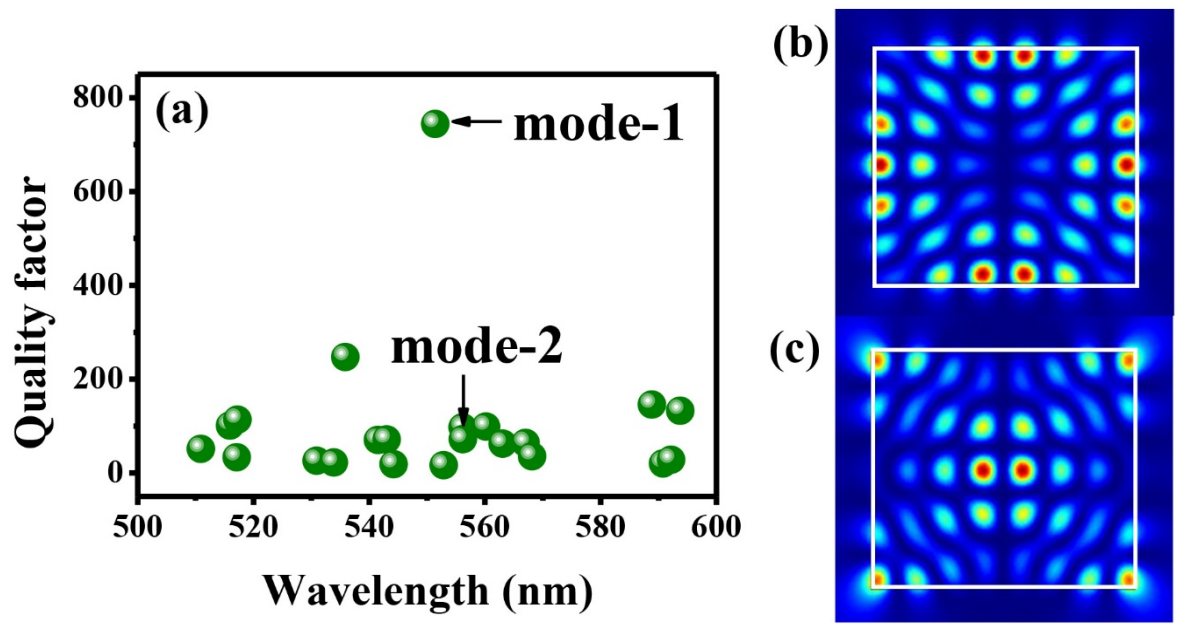

**Figure 3. The resonant modes in the transverse plane.** (a) The Q factors of resonances within the lasing spectral range. (b) The field distributions of mode-1 marked in (a). (c) The field distributions of mode-2 marked in (a).

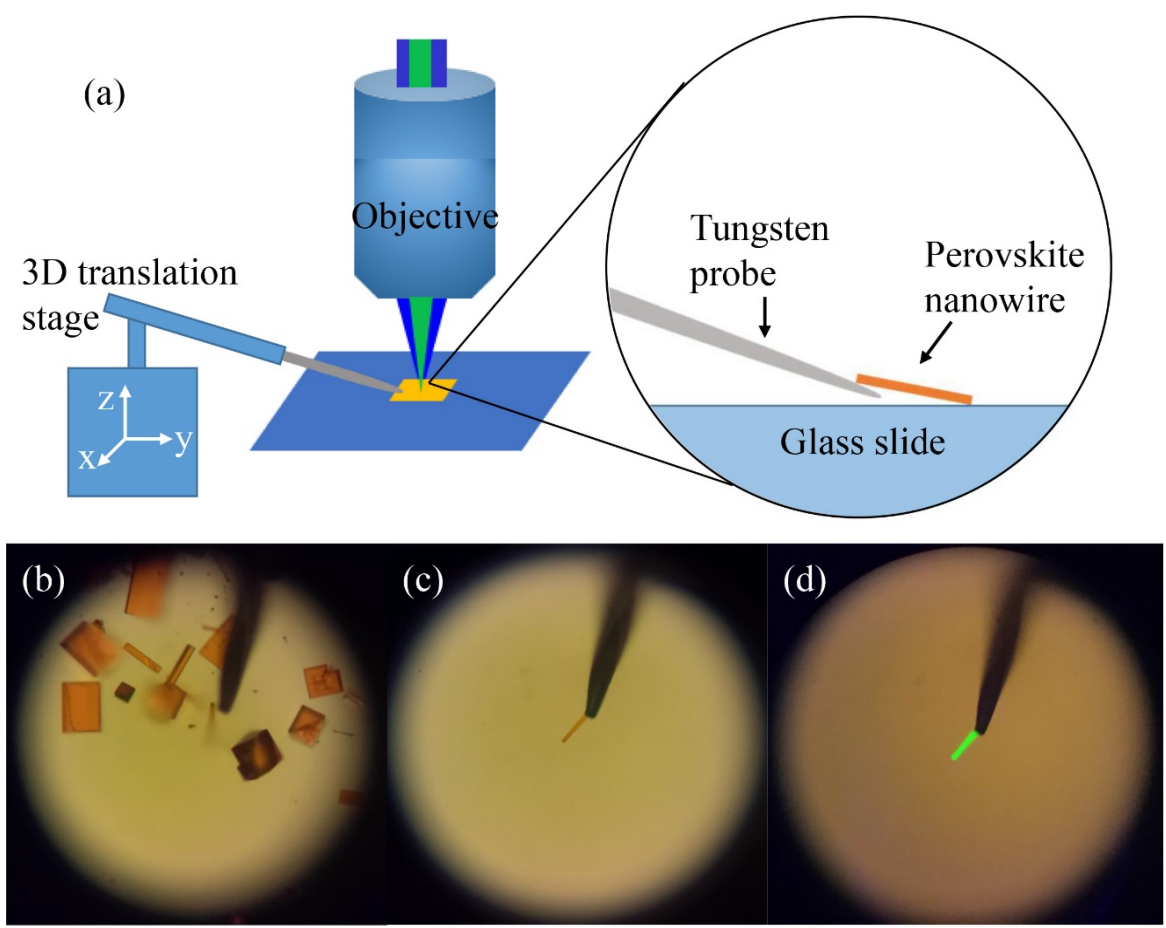

**Figure 4. The nano-manipulation under optical microscope.** (a) The schematic picture of nano-manipulation. (b) The microscope image of selected nanowire. (c) The microscope image of lifted nanowire. (d) The fluorescent microscope image of picked nanowire on the probe.

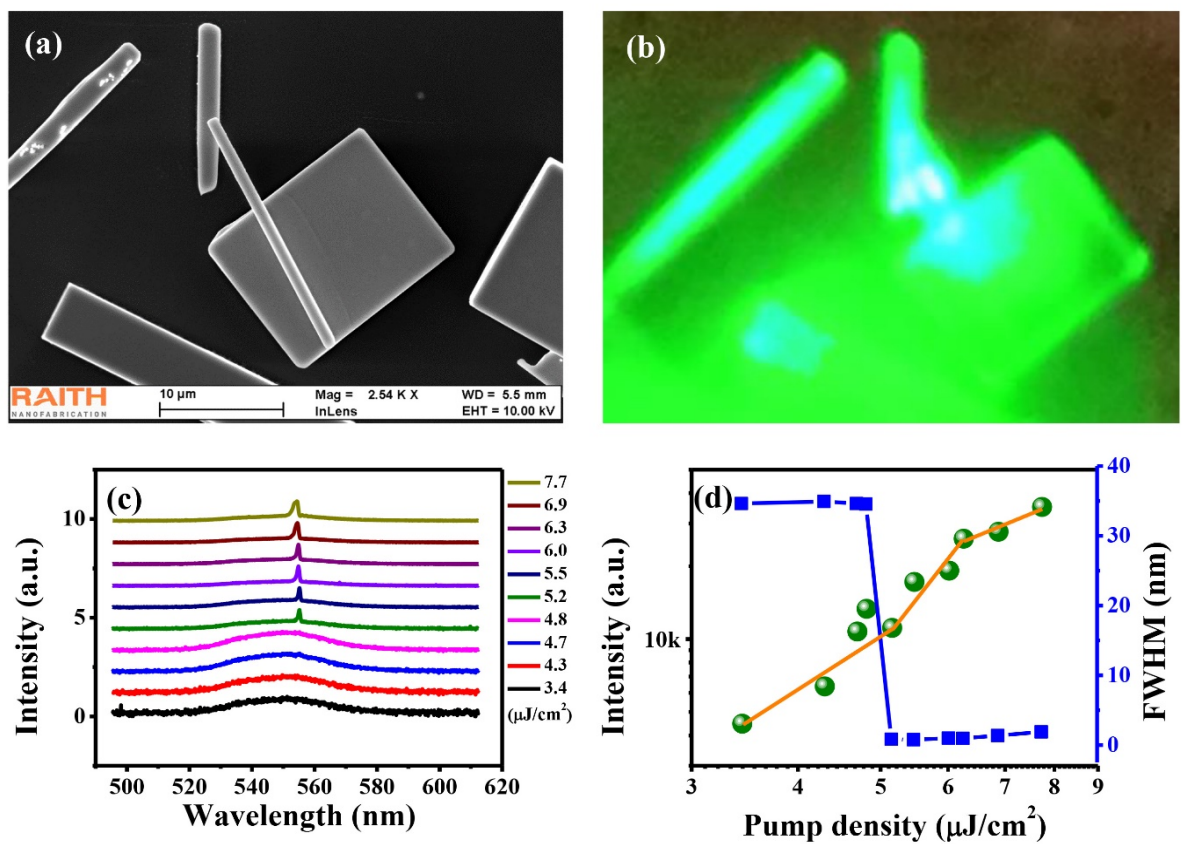

**Figure 5. Single mode lasing action in a perovskite nanowire bridged two nearby perovskite microplates.** (a) SEM image of the structure. (b) Fluorescent image of the nanowire above lasing threshold. Here the pumping density is 7.7 μJ/cm$^2$. (c) Evolution spectra from spontaneous emission to single mode lasing. (d) Integrated emission intensity and FWHM as function of pump density.